\documentclass[11pt,preprint]{aastex}

\newcommand{\etal}{{\it et~al.}}

\begin{document}

\title{Preliminary Analysis of WISE/NEOWISE 3-Band Cryogenic and Post-Cryogenic Observations of Main Belt Asteroids}

\author{Joseph R. Masiero\altaffilmark{1}, A. K. Mainzer\altaffilmark{1}, T. Grav\altaffilmark{2}, J. M. Bauer\altaffilmark{1,3}, R. M. Cutri\altaffilmark{3}, C. Nugent\altaffilmark{4}, M. S. Cabrera\altaffilmark{1,5}}

\altaffiltext{1}{Jet Propulsion Laboratory/California Institute of Technology, 4800 Oak Grove Dr., MS 321-520, Pasadena, CA 91109, USA, {\it Joseph.Masiero@jpl.nasa.gov}}
\altaffiltext{2}{Planetary Science Institute, Tucson, AZ 85719 USA}
\altaffiltext{3}{Infrared Processing and Analysis Center, California Institute of Technology, Pasadena, CA 91125 USA}
\altaffiltext{4}{University of California, Los Angeles, CA, 90095} 
\altaffiltext{5}{California State Polytechnic University Pomona, Pomona, CA, 91768} 

\begin{abstract}

We present preliminary diameters and albedos for 13511 MBAs that were
observed during the 3-Band Cryo phase of the WISE survey (after the
outer cryogen tank was exhausted) and as part of the NEOWISE Post-Cryo
Survey (after the inner cryogen tank was exhausted).  With a reduced
or complete loss of sensitivity in the two long wavelength channels of
WISE, the uncertainty in our fitted diameters and albedos is increased
to $\sim20\%$ for diameter and $\sim40\%$ for albedo.  Diameter fits
using only the $3.4$ and $4.6~\mu$m channels are shown to be dependent
on the literature optical $H$ absolute magnitudes.  These data allow
us to increase the number of size estimates for large MBAs which have
been identified as members of dynamical families.  We present thermal
fits for $14$ asteroids previously identified as the parents of a
dynamical family that were not observed during the fully cryogenic
mission.

\end{abstract}

\section{Introduction}

In \citet[][hereafter Mas11]{masiero11} we presented thermal model
fits for $129,750$ Main Belt asteroids that were observed during the
fully cryogenic portion of the Wide-field Infrared Survey Explorer
\citep[WISE,][]{wright10} mission, which ran from 7 January 2010 to 6
August 2010.  Sensitivity to Solar system objects was enabled by the
NEOWISE augmentation to the WISE mission \citep{mainzer11} which
provided capability for processing and archiving of single-frame
exposures and detection of previously known and new asteroids and
comets.  On 6 August 2010 the hydrogen ice in the outer cryogen tank
was exhausted and the telescope began to warm up, resulting in an
almost immediate loss of the W4 ($22~\mu$m) channel and a decreasing
sensitivity in W3 ($12~\mu$m) beginning the 3-Band Cryo portion of the
mission.  On 29 September 2010 the hydrogen ice in the inner cryogen
reservoir, used to cool the detectors, was exhausted and the W3
channel was lost.  From 29 September 2010 to 1 February 2011, WISE
continued to survey the sky in the NEOWISE Post-Cryo survey phase
\citep{mainzer12pc}, searching for new near-Earth objects (NEOs) and
completing the survey of the largest Main Belt asteroids (MBAs) using
the two shortest bandpasses: W1 ($3.4~\mu$m) and W2 ($4.6~\mu$m).

MBAs have temperatures of $\sim200~$K, depending on their distance
from the Sun and surface properties.  This places the peak of their
blackbody flux near $\lambda_{peak}\sim15~\mu$m.  During the fully
cryogenic portion of the WISE mission the W3 bandpass straddled this
peak and was the primary source of data used for identification and
analysis of the thermal emission from MBAs.  For objects detected
during the 3-Band Cryo portion of the mission we used the W3 data to
constrain the thermal emission, and thus the diameter, of the
objects observed.  As the telescope warmed up, the integration times
in W3 were shortened to prevent saturation of the detectors from the
increasing thermal emission of the telescope \citep{cutri12},
resulting in a decrease in sensitivity to sources in the bandpass.
During the Post-Cryo Survey only W1 and W2 were operational: for MBAs
W1 was sensitive solely to reflected light, while W2 was a blend of
reflected and emitted flux dictated by the object's physical and
orbital parameters (e.g. distance to Sun at the time of observation,
surface temperature, albedo, etc.).

In this work, we present preliminary thermal model fits for 13511 Main
Belt asteroids observed during the 3-Band Cryo phase of the WISE
survey and the NEOWISE Post-Cryo Survey.  During the fully cryogenic
portion of the survey, detectability of most minor planets was
dominated by their thermal emission and so was essentially independent
of their albedo \citep{mainzer11neo}.  However, the Post-Cryo Survey
data at $3.4~\mu$m and $4.6~\mu$m are a mix of reflected and emitted
light.  Thus detectability is strongly coupled to albedo.
Additionally, objects with lower temperatures will have a smaller
thermal emission component to their flux in the W2 band, resulting in
a less accurate estimate of diameter.  In general, diameter fits using
either the 3-Band Cryo or the Post-Cryo Survey data will typically
have larger errors and lower precision than fits from the fully
cryogenic survey given in Mas11, though they still provide useful
information about the observed population of MBAs.

One of the drivers for completing the NEOWISE survey of the inner Main
Belt after the cryogen was exhausted was to have a complete census of
the largest asteroids, particularly those that may be members of
asteroid families.  Having this list allows us to constrain the mass
of the pre-breakup body and more precisely model the age of the family
\citep{vok06,masiero12}.  We present in this work preliminary albedos
and diameters for objects observed during the 3-Band Cryo and
Post-Cryo Survey and discuss the accuracy of these values because
these fits use data processed with the preliminary survey calibration
values.  Future work by the NEOWISE team will include second-pass
processing of the raw data with finalized calibration values as well
as extraction of sources at lower signal-to-noise that will precede a
final release of albedos and diameters.

\section{Observations}

In Mas11 we focused our analysis on data taken during the fully
cryogenic portion of the WISE mission.  For this work, we analyze the
3-Band Cryo and Post-Cryo Survey observations taken by WISE as part of
the NEOWISE survey.  Observations obtained between Modified Julian
Dates (MJDs) of 55414 and 55468 are available in the 3-Band Cryo
Single-Exposure database, served by the Infrared Science Archive
(IRSA)\footnote{\it http://irsa.ipac.caltech.edu}.  Post-Cryo data,
spanning a MJD range of 55468 to 55593, are archived in the NEOWISE
Preliminary Post-Cryo database and also served by IRSA.  Data from the
3-Band Cryo survey were released to the public on 29 June
2012\footnote{\it http://wise2.ipac.caltech.edu/docs/release/3band/}
and preliminary data from the NEOWISE Post-Cryo Survey were released
to the public on 31 July 2012\footnote{\it
  http://wise2.ipac.caltech.edu/docs/release/postcryo\_prelim/}.  We
note that the Post-Cryo Survey data have only undergone first-pass
processing, and users are strongly encouraged to consult the
Explanatory Supplement \citep{cutri12} associated with the database.

We follow the same method as described in Mas11 to acquire
detections of MBAs that have been vetted both by our internal WISE
Moving Object Processing System \citep[WMOPS;][]{mainzer11} and by the
Minor Planet Center (MPC).  This includes the use of the same quality
flag settings from the pipeline extraction for cleaning of detections
before thermal fitting as discussed in Mas11.  Of the 14638 objects
observed by WISE between MJDs 55414 and 55593, 13511 MBAs had data of
sufficient quality to perform thermal model fits.

Due to the nature of WISE's orbit and the synodic period of MBAs,
approximately half of the objects observed during the 3-Band Cryo and
Post-Cryo Survey had also been observed earlier during the fully
cryogenic survey.  We use these overlap objects as standards to
evaluate the accuracy of the thermal model fits using these data (see
Section~\ref{sec.overlap}).  While in some cases extremely irregularly
shaped slow-rotating objects may show significant changes in projected
area between epochs and thus large variations in both emitted and
reflected flux, this is expected to be a small fraction of all objects
observed and only to add a small component of random error to the
comparison \citep{grav11troj}.

\section{Thermal Fitting}
\label{sec.models}
Following the procedure discussed in Mas11, we use a faceted NEATM
thermal model to determine the diameter and albedo of the MBAs
observed after the outer cryogen tank was exhausted.  In most cases we
only have thermal emission data in a single band, and so we are forced
to assume a beaming parameter for the models.  We use a beaming
parameter of $\eta=1.0\pm0.2$, based on the peak of the distribution
for MBAs given in Mas11.  Our measured flux in W2 is typically
dominated by thermal emission, however the reflected component of the
W2 flux will influence our models.

In order to remove the reflected component from the measured W2 flux,
we need to determine the optical geometric albedo ($p_V$) and assume a
ratio between the near-IR (NIR) and optical albedos.  In Mas11 we
were able to fit this ratio for objects with with observations in $W3$
and/or $W4$ as well as $W1$ and $W2$, however we cannot do this for
the Post-Cryo Survey data.  Following the best-fit value from Mas11,
for those objects we assume a NIR/optical reflectance ratio of
$1.4\pm0.5$.  In all cases, we also assume that the reflectivity in W1
is identical to that in W2 ($p_{W1}=p_{W2}=p_{NIR}$).  For objects
with very red spectral slopes this may not necessarily be a good
assumption \citep[cf.][]{mainzer11tax, grav12trojtax} however without
additional data (e.g. spectral taxonomy) it is impossible to
disentangle these two values for this dataset.

To determine optical albedo we used the $H$ absolute magnitude and $G$
slope parameter given in the Minor Planet Center's MPCORB
file\footnote{\it http://www.minorplanetcenter.net/iau/MPCORB.html},
and updated using other databases following Mas11.  We note that
recent work has shown that these $H$ values may be systematically
offset in some magnitude ranges by up to $0.4~$mags when comparing
predicted and observed apparent magnitudes \citep{pravec12}.  This
will affect the albedos that we calculate for the asteroids presented
here, which in turn will change the relative contribution of emitted
and reflected light in W2.  Unlike the results presented in Mas11,
where the diameter determination is independent of the optical $H$
measurement, any future revision to the measured $H$ values will
require a refitting of the thermal models and will likely result in an
change in modeled diameter.

\section{Discussion}
\subsection{Comparison of Overlap Objects}
\label{sec.overlap}

\citet{mainzer12pc} showed a comparison between the thermal fits
performed with the Post-Cryo Survey data and non-radiometrically
determined diameters for a range of NEOs and MBAs to derive a relative
accuracy of $\sim20\%$ on diameter and $\sim40\%$ on albedo.  As a
parallel check we have taken objects that were observed both before
and after the exhaustion of the outer cryogen reservoir and compared
the diameters and albedos found here to those values given in Mas11.
Of the fits presented here, 7222 unique objects also appeared in the
fully cryogenic observations that were presented in Mas11.  Of
these, 2844 were observed during the 3-Band Cryo phase of the survey
and 4403 were observed during the Post-Cryo Survey (note that 25
objects appeared in all three phases of the survey).

\begin{figure}[ht]
\begin{center}
\includegraphics[scale=0.5]{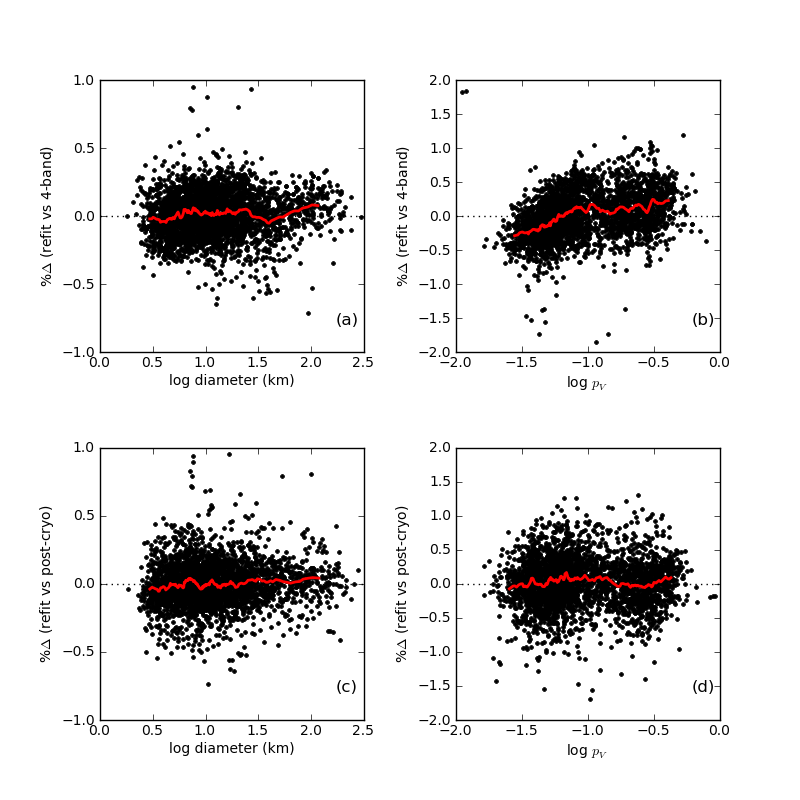}
\protect\caption{Comparison of thermal fits for objects appearing in
  both the fully cryogenic data set as well as the Post-Cryo Survey
  data.  The left column shows the comparison of the diameters
  ($\log~D$) while the right column shows the comparison of the visual
  albedos ($\log~p_V$).  The top row shows the fractional difference
  between the 4-band fits presented in Mas11 and refits of those
  data using only the W1 and W2 bandpasses, while the bottom row shows
  the fractional difference between the 2-band refit of the fully
  cryogenic data and the 2-band fits of the Post-Cryo Survey data.
  The dotted line in each case shows a one-to-one relationship, and
  the solid red line shows a running box average for each comparison.}
\label{fig.refit}
\end{center}
\end{figure}

For all objects seen in both the Post-Cryo Survey and in the fully
cryogenic 4-band survey, we have refit the 4-band Cryo data using only
the W1 and W2 measurements as a way of differentiating changes in the
quality of fit due to the loss of W3 and W4 sensitivity from changes
due to the different observing circumstances.  The results of this
test are shown in Figure~\ref{fig.refit}.  We include a running box
average of the data in order to assess the population trends, which
bins by 100 objects, in steps of 20.  In general these tests follow
the expected one-to-one relationship, with the exception of the
comparison of the 2-band and 4-band fits of the fully cryogenic data
(Figure~\ref{fig.refit}b), which deviates at both high and low
albedos, and effect that was also observed for the NEOs in the
Post-Cryo Survey data by \citet{mainzer12pc}.  \citep{mainzer11tax}
have shown that high albedo objects tend to have optical/NIR
reflectance ratios of $\sim1.6$, while objects with low albedos tend
to have reflectance ratios of $\sim1.0$ (though D-type objects deviate
from this trend and have very large reflectance ratios).  As we use a
fixed reflectance ratio of $1.4$, low albedo objects with W1
measurements will have a final fitted $p_V$ below the true value,
while high albedo objects will have a $p_V$ slightly above, which
corresponds to the twist observed in Figure~\ref{fig.refit}b.

Comparison of the 2-band refits to the results from Mas11 show the
uncertainty induced by the loss of W3 and W4 information
(Figure~\ref{fig.refit}a-b) results in a $1\sigma$ scatter of $16\%$
in diameter and $32\%$ in albedo (three points fall outside the
plotted range for Figure~\ref{fig.refit}(a); all other panels show all
objects considered).  Comparison of the 2-band refits of the fully
cryogenic data to the Post-Cryo Survey fits
(Figure~\ref{fig.refit}c-d) shows the errors induced by both changes
in observing aspect as well as calibration differences between the two
data sets, which collectively result in a $1\sigma$ scatter of $13\%$
in diameter and $31\%$ in albedo as well.  Combined, these two errors
result in a measured $21\%$ relative error on diameter and $45\%$ in
albedo.

These total errors are in line with what was found by
\citet{mainzer12pc} when comparing the fits from Post-Cryo Survey data
to literature diameters.  Our errors are also in line with the
uncertainties measured for the ExploreNEOs project which uses a
similar pair of bandpasses ($3.6~\mu$m and $4.5~\mu$m) from the Warm
{\it Spitzer} mission to model diameters and albedos for previously
known NEOs \citep{trilling10,harris11}.  Our measured level of error
indicates that the random error introduced by the combined effect of
irregular shape and observing geometry is below this level.  We note
that the method of source extraction used for all phases of the WISE
data processing relies on the position of the object in all detected
bands.  In general the W1 and W2 measurements from the 4-band data
will be at lower signal-to-noise than the data for those objects
extracted from the Post-Cryo Data, inflating the errors quoted above.

\begin{figure}[ht]
\begin{center}
\includegraphics[scale=0.5]{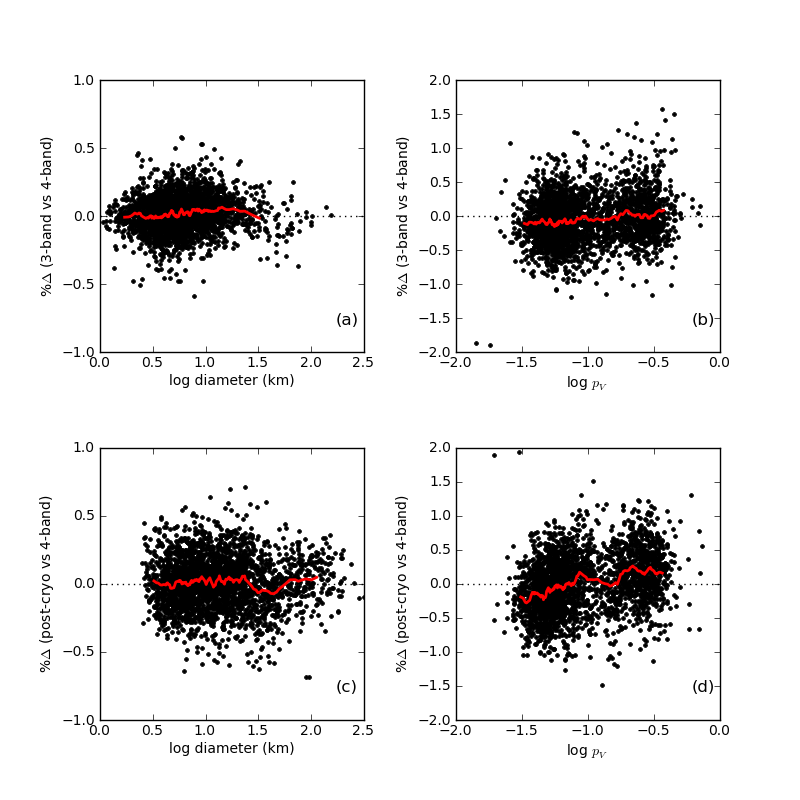}
\protect\caption{Comparison of thermal fits for objects appearing both
  in the fully cryogenic data set as well as in the 3-Band Cryo or
  Post-Cryo Survey data.  The left column shows the comparison of the
  diameters ($\log~D$) while the right column shows the comparison of
  the visual albedos ($\log~p_V$).  The top row shows the fractional
  difference between the 3-Band Cryo fits and the values from Mas11,
  while the bottom row shows the fractional difference between the
  2-band Post-Cryo Survey fits and the Mas11 values.  The dotted
  line in each case shows a one-to-one relationship, and the solid red
  line shows a running box average for each comparison.}
\label{fig.overlap}
\end{center}
\end{figure}

We show in Figure~\ref{fig.overlap} the comparison between the fits
for objects appearing in the fully cryogenic data as well as the
3-Band Cryo or Post-Cryo Survey data.  As in Figure~\ref{fig.refit} we
include a running box average using the same parameters as above.  We
see no large-scale systematic shifts between datasets, however we do
confirm the increase in scatter in the fits using the latter data
sets.  We note that in Figure~\ref{fig.overlap}d shows a behavior
similar to what we observe in Figure~\ref{fig.refit}b, where the fits
of albedo deviate to more extreme values for both high and low albedo
objects.  As discussed above, this is attributed to the use of an
assumed optical/NIR reflectance ratio that is between the values
measured for high and low albedo objects when they are considered
independently.

\subsection{Preliminary Diameters and Albedos}

\begin{figure}[ht]
\begin{center}
\includegraphics[scale=0.5]{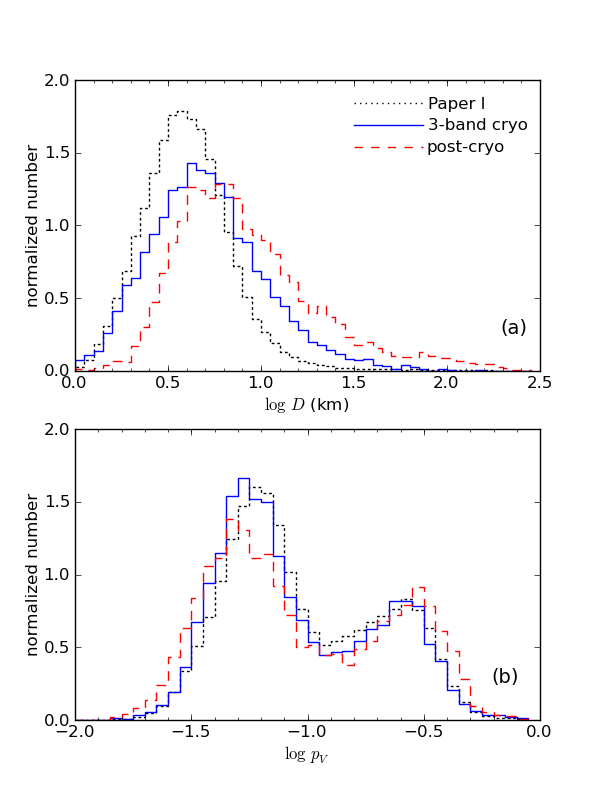}
\protect\caption{Preliminary diameter (a) and albedo (b) distributions
  for all MBAs from Mas11 (black dotted), MBAs from the 3-Band Cryo
  data (blue solid), and MBAs from the Post-Cryo Survey (red dashed).
  Note the scales are normalized: the total number of objects
  presented in Mas11 is over an order of magnitude larger than the
  other two populations.}
\label{fig.fithist}
\end{center}
\end{figure}

We present in Table~\ref{tab.pcFits} the preliminary fitted diameters
and albedos for all MBAs observed during the 3-Band Cryo and Post-Cryo
surveys, along with their associated errors (note that errors do not
include the systematic $\sim20\%$ diameter error or $\sim40\%$ albedo
error discussed above).  We also include the number of detections used
in each band as well as the $H$ and $G$ values used for the fit.
Objects without measured visible magnitudes have ``nan'' entered for
their $H$, $G$, and albedo values.  The recommended method for
extracting fluxes for asteroid detections is discussed in
\citet{mainzer11} and \citet{cutri12}.  Figure~\ref{fig.fithist} shows
the preliminary diameter and albedo distributions for the asteroids
observed during 3-Band Cryo and Post-Cryo Surveys compared to the
population presented in Mas11.  With the loss of the long wavelength
channels the sensitivity to small objects was reduced and peak of the
diameter distribution moves to larger sizes.  In both the 3-Band Cryo
and Post-Cryo Survey data we see a shift in the high and low branches
of the albedo distribution to more extreme values when compared to the
population from Mas11.  This shift was also observed for the NEOs by
\citet{mainzer12pc}, and attributed to the forced values for both
beaming and NIR/optical reflectance ratio in the fits of the Post-Cryo
Survey data.

Reflected light is a much more significant component in the W1 and W2
bandpasses for MBAs than in the W3 and W4 bands used in Mas11 to
perform thermal fits.  As such, results from the model fits presented
here are inherently tied to the optical measurements and cannot be
considered insensitive to albedo as was assumed in Mas11.  This bias
will most strongly affect objects that are small and have low albedos.
Thus, care must be taken before extrapolating the trends observed in
the these fits to the greater MBA population.

\begin{table}[ht]
\begin{center}
\caption{Thermal model fits for MBAs in the 3-Band Cryo and NEOWISE
  Post-Cryo Survey.  Table 1 is published in its entirety in the
  electronic edition of ApJL; a portion is shown here for guidance
  regarding its form and content.}
\vspace{1ex}
\noindent
\begin{tabular}{ccccccccc}
\tableline
  Name  &   H   &    G  &       D (km)       &        $p_V$      & n$_{W1}$ & n$_{W2}$ & n$_{W3}$ \\ 
\tableline
  00003 &  5.33 &  0.32 & 246.60 $\pm$ 10.59 & 0.214 $\pm$ 0.026 & 11 & 11 &  0 \\ 
  00005 &  6.85 &  0.15 & 106.70 $\pm$  3.14 & 0.282 $\pm$ 0.050 & 14 & 14 &  0 \\ 
  00011 &  6.55 &  0.15 & 154.13 $\pm$  3.92 & 0.178 $\pm$ 0.030 & 12 & 12 &  0 \\ 
  00014 &  6.30 &  0.15 & 145.68 $\pm$  5.27 & 0.251 $\pm$ 0.041 &  9 &  9 &  0 \\ 
  00016 &  5.90 &  0.20 & 288.29 $\pm$  4.63 & 0.093 $\pm$ 0.024 &  5 &  5 &  0 \\ 
  00017 &  7.76 &  0.15 &  69.64 $\pm$  2.26 & 0.287 $\pm$ 0.051 &  9 & 10 &  0 \\ 
  00018 &  6.51 &  0.25 & 155.84 $\pm$  5.63 & 0.181 $\pm$ 0.033 & 12 & 12 &  0 \\ 
  00019 &  7.13 &  0.10 & 209.81 $\pm$  2.20 & 0.056 $\pm$ 0.012 & 11 & 11 &  0 \\ 
  00020 &  6.50 &  0.25 & 135.68 $\pm$  3.67 & 0.241 $\pm$ 0.018 & 13 & 13 &  0 \\ 
\hline
\end{tabular}
\label{tab.pcFits}
\end{center}
\end{table}

\subsection{Asteroid Family Members}

One of the primary drivers of the Post-Cryo Survey was to complete the
census of large MBAs that are related to dynamically associated
asteroid families.  The largest body in a family anchors both the mass
estimate of the pre-breakup body and the starting point for family age
simulations.  The 3-Band Cryo and Post-Cryo Survey data contain 3319
objects identified by \citep{nesvornyPDS} as members of asteroid
families that were able to have thermal models fit to their
measurements.  Of these, 14 were identified as family parents and were
not observed during the WISE fully cryogenic mission, including (3)
Juno, (20) Massalia, (44) Nysa, (170) Maria, (298) Baptistina, (363)
Padua, (434) Hungaria, (490) Veritas, (569) Misa, (778) Theobalda,
(1270) Datura, (1892) Lucienne, (4652) Iannini, and (7353) Kazuya.

Of these 14 bodies that are the largest in their family, only 4 had
albedos below $p_V=0.1$, in contrast with the general population
presented here where $\sim60\%$ of the MBAs had low albedos.  This is
due to a number of overlapping selection biases, including the
dominance of high albedo objects in the literature family lists
(cf. Mas11), preferential sensitivity to high albedo objects in the
3-Band Cryo and Post-Cryo Survey data compared to Mas11 (meaning
this data set is more likely to miss low albedo asteroids), and the
longer synodic periods of MBAs with smaller semimajor axes.  The
differences in synodic period resulted in a larger fraction of objects
in the inner Main Belt that were not observed during the fully
cryogenic phase of the WISE survey, compared to the outer Main Belt.
Future work in family identification will begin to mitigate these
biases.

\section{Conclusions}

We present preliminary thermal model fits for 13511 MBAs using
observations acquired by the WISE and NEOWISE surveys following the
exhaustion of the outer cryogen tank that marked the end of the fully
cryogenic WISE survey.  Accuracy of these fits is degraded with
respect to the results discussed in Mas11 due to the loss of the W3
and W4 bandpasses, however fits of diameter with relative accuracy of
$\sim20\%$ are still possible.  Unlike the fits presented in Mas11,
these determinations depend strongly on the measured value of the
optical albedo (as calculated from the $H$ absolute magnitude).  Thus,
any revision to the $H$ values will require a new thermal model to be
fit to the data.  This dataset includes detection of 3319 members of
previously identified asteroid families, one of the main goals of the
Post-Cryo Survey.  Future work by the NEOWISE team will include
second-pass processing of these data sets using updated calibration
products, which is expected to improve the accuracy of diameter and
albedo determination.

\section*{Acknowledgments}

J.R.M. was supported by an appointment to the NASA Postdoctoral
Program at JPL, administered by Oak Ridge Associated Universities
through a contract with NASA.  We thank the anonymous referee for
their helpful comments.  This publication makes use of data products
from the Wide-field Infrared Survey Explorer, which is a joint project
of the University of California, Los Angeles, and the Jet Propulsion
Laboratory/California Institute of Technology, funded by the National
Aeronautics and Space Administration.  This publication also makes use
of data products from NEOWISE, which is a project of the Jet
Propulsion Laboratory/California Institute of Technology, funded by
the Planetary Science Division of the National Aeronautics and Space
Administration.  This research has made use of the NASA/IPAC Infrared
Science Archive, which is operated by the Jet Propulsion Laboratory,
California Institute of Technology, under contract with the National
Aeronautics and Space Administration.


\clearpage
\begin{thebibliography}{XXX}

\bibitem[Cutri \etal(2012)]{cutri12}
Cutri, R.M., Wright, E.L., Conrow, T., Bauer, J., \etal, 2012, {\it http://wise2.ipac.caltech.edu/docs/release/allsky/expsup/index.html}

\bibitem[Grav \etal(2011)]{grav11troj}
Grav, T., Mainzer, A.K., Bauer, J.M., Masiero, J., Spahr, T, \etal, 2011, ApJ, 742, 40.

\bibitem[Grav \etal(2012)]{grav12trojtax}
Grav, T., Mainzer, A.K., Bauer, J.M., Masiero, J., 2012, ApJ in press.

\bibitem[Harris \etal(2011)]{harris11}
Harris, A.W., Mommert, M., Hora, J.L., Mueller, M., \etal, 2011, AJ, 141, 75.

\bibitem[Mainzer \etal(2011a)]{mainzer11}
Mainzer, A.K., Bauer, J.M., Grav, T., Masiero, J., \etal, 2011a, ApJ, 731, 53.

\bibitem[Mainzer \etal(2011b)]{mainzer11tax}
Mainzer, A.K., Grav, T., Masiero, J., Hand, E., \etal, 2011b, ApJ, 741, 90.

\bibitem[Mainzer \etal(2011c)]{mainzer11neo}
Mainzer, A.K., Grav, T., Bauer, J.M., Masiero, J., \etal, 2011c, ApJ, 743, 156.

\bibitem[Mainzer \etal(2012)]{mainzer12pc}
Mainzer, A.K., Grav, T., Bauer, J.M., Masiero, J., \etal, 2012, ApJ, submitted to ApJL.

\bibitem[Masiero \etal(2011)]{masiero11}
Masiero, J.R, Mainzer, A.K., Grav, T., Bauer, J.M., \etal, 2011, ApJ, 741, 68.

\bibitem[Masiero \etal(2012)]{masiero12}
Masiero, J.R, Mainzer, A.K., Grav, T., Bauer, J.M. \& Jedicke, R., 2012, ApJ in press.

\bibitem[Nesvorn\'{y}(2010)]{nesvornyPDS}
Nesvorny, D., 2010, EAR-A-VARGBDET-5-NESVORNYFAM-V1.0, NASA Planetary Data System.

\bibitem[Pravec \etal(2012)]{pravec12}
Pravec, P., Harris, A.W., Ku\u{s}nir\'{a}k, P., Gal\'{a}d, A. \& Hornoch, K., 2012, Icarus, submitted.

\bibitem[Trilling \etal(2010)]{trilling10}
Trilling, D., Mueller, M., Hora, J.L., Harris, A.W., \etal, 2012, AJ, 140, 770.

\bibitem[Vokrouhlick\'{y} \etal(2006)]{vok06}
Vokrouhlick\'{y}, D., Bro\u{z}, M., Bottke, W.F., Nesvorn\'{y}, D. \& Morbidelli, A., 2006, Icarus 182, 118.

\bibitem[Wright \etal(2010)]{wright10}
Wright, E.L., Eisenhardt, P., Mainzer, A.K., Ressler, M.E., Cutri, R.M., Jarrett, T., Kirkpatrick, J.D., Padgett, D., \etal, 2010, AJ, 140, 1868.


\end{thebibliography}
\end{document}